\documentclass[%
 reprint,
 amsmath,amssymb,
 aps,
prl
]{revtex4-1}
\pdfoutput=1
\usepackage{graphicx}
\usepackage{dcolumn}
\usepackage{bm}
\usepackage{subfigure}

\begin{document}
\preprint{APS/123-QED}

\title{Erratum for ``Axion Dark Matter Coupling to Resonant Photons via Magnetic Field"}
\author{Ben T. McAllister}
\email{ben.mcallister@uwa.edu.au}
\author{Stephen R. Parker}
\affiliation{ARC Centre of Excellence for Engineered Quantum Systems, School of Physics, The University of Western Australia, 35 Stirling Highway, Crawley 6009, Western Australia, Australia}
\author{Michael E. Tobar}
\email{michael.tobar@uwa.edu.au}
\affiliation{ARC Centre of Excellence for Engineered Quantum Systems, School of Physics, The University of Western Australia, 35 Stirling Highway, Crawley 6009, Western Australia, Australia}
\date{\today}%
\maketitle
\begin{figure*}[t]
	\centering
	\includegraphics[scale=0.65]{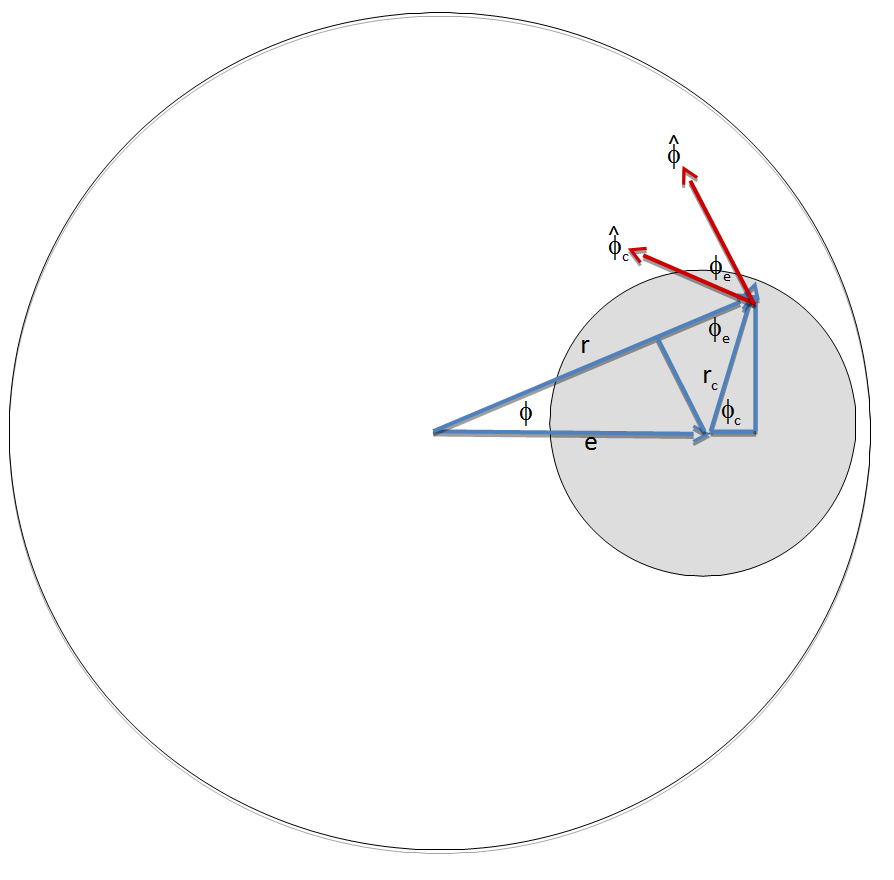}
	\caption{A diagram of the offset cavity experiment, the unit vectors are shown in red. The angle $\phi_e$ is defined as the angle between these unit vectors. All important parameters are labelled. The cavity is shown in grey, and solenoid is shown in white.}
	\label{fig:diagram}
\end{figure*}

We the authors of a recent letter~\cite{OurPRL} wish to present the following erratum detailing an error in our analysis and an error in a recent comment on our analysis~\cite{comment}. In our letter we derived for the first time the magnetic form factor, which gives a general formula for the magnetic coupling of a photon produced via the inverse Primakoff effect in a Sikivie Haloscope axion dark matter detector. Previously only the electric coupling had been considered as, in many cases the two couplings are equal. In general this is not true. In our letter~\cite{OurPRL}, Equations (1) - (16) are all correct, and the general electromagnetic form factor can more completely be written as,
\begin{align*}
	\text{C}_{\text{EM}}&=\frac{C_E+C_B}{2}\\
	&=\frac{\left|\int dV_{c}\vec{E_c}\cdot\vec{\hat z}\right|^2}{2~V_c\int dV_{c}\epsilon_r\mid E_c\mid^2}+\frac{\frac{\omega_a^2}{c^2}\left|\int dV_{c}\frac{r}{2}\vec{B_c}\cdot\vec{\hat\phi}\right|^2}{2~V_c\int dV_{c}\frac{1}{\mu_r}\mid B_c\mid^2}
\end{align*}
Here the values $\mu_r$ and $\epsilon_r$ generalise the expression if dielectric or magnetic materials are present in the Halocope resonator. This is a very important generalisation of the electromagnetic form factor, which has been ignored in the past. For a TM mode in an empty cylindrical cavity, the magnetic form factor can be written as
\begin{align}
	\centering
	\text{C}_{\text{B}}=\frac{\frac{\omega_a^2}{c^2}\left|\int dV_{c}(\text{B}_\text{c}(\text{r}_c)\hat{\phi}_c)\cdot(\frac{r}{2}\hat{\phi})\right|^2}{V\int dV_{c}\mid B_c\mid^2}.\label{eq:CBold}
\end{align}
The cavity magnetic field is in the cavity $\phi_c$ direction (and a function of the cavity radius), whilst the axion induced magnetic field is in the solenoid's $\phi$ direction, and proportional to the radial distance in the solenoid to the point of integration. Generally speaking these two directions are not the same, and in the case for a cavity offset by some distance, $e$, from the centre of the solenoid the dot product is non-trivial, which is the structure analyzed in our letter. Equation (17) in our letter presents a result for this structure, which we show is in error by a factor $r/r_c$; this changes the conclusion. We revisited this calculation because in a recent comment~\cite{comment} on the letter in question the dot product between $\phi_c$ and $\phi$ is claimed to be;
\begin{align}
	\begin{split}
	\hat{\phi}\cdot\hat{\phi}_c&=\cos{(\phi+\phi_c)}\\
	&=\cos{\phi_c}\cos{\phi}-\sin{\phi_c}\sin{\phi}.
\end{split}
	\label{eq:comment}
\end{align}
We show this calculation to be inaccurate. Fig.~\ref{fig:diagram} shows the geometry of the problem under consideration, and we can derive the following expressions from trigonometry;
\begin{align*}
	\cos{\phi}&=\frac{e+r_c\cos{\phi_c}}{r}\\
	\sin{\phi}&=\frac{r_c\sin{\phi_c}}{r}.
\end{align*}
If we follow through with the expression for $\hat{\phi}\cdot\hat{\phi}_c$ in Eq.~\ref{eq:comment}, using the above expressions for $\cos{\phi}$ and $\sin{\phi}$ we arrive at the following expression for $C_B$
\begin{align*}
	\text{C}_{\text{B}}=\frac{\frac{\omega_a^2}{c^2}\left|\int dV_{c}\text{B}_{\text{c}_\phi}\frac{\text{r}_c\cos{2\phi_c}+\text{e}\cos{\phi_c}}{2}\right|^2}{V\int dV_{c}\mid B_c\mid^2},
\end{align*}
which evaluates to zero for all offset values, when integrated over the full cavity limits. Clearly this result is in error, as the form factor is known to be non-zero.
However, careful reanalysis of the system has led us to uncover a minor discrepancy in Equation (17) in our letter, which we will now discuss. In fig.~\ref{fig:diagram} we define the angle between the two unit vectors as $\phi_e$. Since both are unit vectors (of magnitude one) the dot product can be written as
\begin{align*}
\hat{\phi}\cdot\hat{\phi}_c&=\cos{\phi_e}.
\end{align*}
Thus Eq.~\ref{eq:CBold} becomes
\begin{align*}
\text{C}_\text{B}&=\frac{\frac{\omega_a^2}{c^2}\left|\int dV_{c}\text{B}_\text{c}(\text{r}_c)\frac{r}{2}\cos{(\phi_e)}\right|^2}{V\int dV_{c}\mid B_c\mid^2}.
\end{align*}
where
\begin{align*}
\cos{\phi_e}&=\frac{r-e\cos{\phi}}{r_c},
\end{align*}
(which comes from trigonometry as can be seen in fig.~\ref{fig:diagram}). Thus, the integral reduces to
\begin{equation}
\text{C}_{\text{B}}=\frac{\frac{\omega_a^2}{c^2}\left|\int dV_{c}\text{B}_{\text{c}_\phi}\frac{\text{r}-\text{e}\cos{\phi}}{2}\times\frac{r}{r_c}\right|^2}{V\int dV_{c}\mid B_c\mid^2}.\label{eq:correctCB}
\end{equation}
This expression differs to Equation (17) in the original letter by a factor of $r/r_c$. At some point in our analysis, a factor of $r/r_c$ has propagated through our numerical calculations, causing an error in the presented results. It is important to note that when Eq.~\ref{eq:correctCB} is evaluated over the limits of the cavity we find that it is constant with varying offset, and equal to $C_E$ for TM modes in empty cylindrical cavity resonators.  Despite the form factor remaining constant as the position of the cavity changes within the solenoid for this example, we maintain that the magnetic coupling is an extremely important parameter in analysis of generalized haloscope systems, as there is no a priori reason to expect that the electric and magnetic couplings are always equal in general. Particularly, systems which introduce dielectrics (or magnetic materials) into the cavity must consider this in estimates of their sensitivity, and systems with spatially separated electric and magnetic fields may not be accurately analyzed without consideration of the magnetic coupling. In fact, a full analysis of such a system is the topic of a proposal regarding the use of lumped 3D LC resonators in axion detection~\cite{arXivProposal}.\\This work was supported by Australian Research Council grants CE110001013, as well as the Australian Postgraduate Award and the Bruce and Betty Green Foundation.\\


\newpage
\begin{thebibliography}{28}%
		\makeatletter
		\providecommand \@ifxundefined [1]{%
			\@ifx{#1\undefined}
		}%
		\providecommand \@ifnum [1]{%
			\ifnum #1\expandafter \@firstoftwo
			\else \expandafter \@secondoftwo
			\fi
		}%
		\providecommand \@ifx [1]{%
			\ifx #1\expandafter \@firstoftwo
			\else \expandafter \@secondoftwo
			\fi
		}%
		\providecommand \natexlab [1]{#1}%
		\providecommand \enquote  [1]{``#1''}%
		\providecommand \bibnamefont  [1]{#1}%
		\providecommand \bibfnamefont [1]{#1}%
		\providecommand \citenamefont [1]{#1}%
		\providecommand \href@noop [0]{\@secondoftwo}%
		\providecommand \href [0]{\begingroup \@sanitize@url \@href}%
		\providecommand \@href[1]{\@@startlink{#1}\@@href}%
		\providecommand \@@href[1]{\endgroup#1\@@endlink}%
		\providecommand \@sanitize@url [0]{\catcode `\\12\catcode `\$12\catcode
			`\&12\catcode `\#12\catcode `\^12\catcode `\_12\catcode `\%12\relax}%
		\providecommand \@@startlink[1]{}%
		\providecommand \@@endlink[0]{}%
		\providecommand \url  [0]{\begingroup\@sanitize@url \@url }%
		\providecommand \@url [1]{\endgroup\@href {#1}{\urlprefix }}%
		\providecommand \urlprefix  [0]{URL }%
		\providecommand \Eprint [0]{\href }%
		\providecommand \doibase [0]{http://dx.doi.org/}%
		\providecommand \selectlanguage [0]{\@gobble}%
		\providecommand \bibinfo  [0]{\@secondoftwo}%
		\providecommand \bibfield  [0]{\@secondoftwo}%
		\providecommand \translation [1]{[#1]}%
		\providecommand \BibitemOpen [0]{}%
		\providecommand \bibitemStop [0]{}%
		\providecommand \bibitemNoStop [0]{.\EOS\space}%
		\providecommand \EOS [0]{\spacefactor3000\relax}%
		\providecommand \BibitemShut  [1]{\csname bibitem#1\endcsname}%
		\let\auto@bib@innerbib\@empty
\bibitem{OurPRL}
Ben~T. McAllister, Stephen~R. Parker, and Michael~E. Tobar.
\newblock Axion dark matter coupling to resonant photons via magnetic field.
\newblock {\em Phys. Rev. Lett.}, 116:161804, Apr 2016.
\bibitem{comment}
		Sangjun Lee, Sung~Woo Youn, and Y.~K. Semertzidis.
		\newblock {Comment on``Axion Dark Matter Coupling to Resonant Photons via
			Magnetic Field"}.
		\newblock arXiv:1606.09504
		\newblock (2016).
\bibitem{arXivProposal}
Ben~T. McAllister, Stephen~R. Parker, and Michael~E. Tobar.
\newblock 3D lumped LC resonators as low mass axion haloscopes
\newblock arXiv:1605.05427 [physics.ins-det]
\newblock (2016).

\end{thebibliography}
\end{document}